\documentclass[preprint,showpacs,showkeys,superscriptaddress]{revtex4}
\usepackage{isolatin1,graphicx,amsmath}
\usepackage[hypertex]{hyperref}

\newcommand{\average}[1]{\left\langle{#1}\right\rangle}

\newcommand{\expj}[1]{\mathrm{E_J}\left[{#1}\right]}

\begin{document}

\title{A note on the Guerra and Talagrand theorems for Mean Field Spin
  Glasses: the simple case of spherical models}
\author{Silvio Franz and Francesca Tria}

  \affiliation{{\it  The Abdus Salam International Center for
Theoretical Physics}\\
  {\it Strada Costiera 11, P.O. Box 563,
    I-34100 Trieste (Italy)}\\
  {\it E-mail: franz@ictp.trieste.it, ftria@ictp.trieste.it}}

\date{\today}

\begin{abstract}
  The aim of this paper is to discuss the main ideas of the Talagrand
  proof of the Parisi Ansatz for the free-energy of Mean Field Spin
  Glasses with a physicist's approach.  We consider the case of the
  spherical $p$-spin model, which has the following advantages: 1) the
  Parisi Ansatz takes the simple ``one step replica symmetry breaking
  form'', 2) the replica free-energy as a function of the order
  parameters is simple enough to allow for numerical maximization with
  arbitrary precision.  We present the essential ideas of the proof,
  we stress its connections with the theory of effective potentials
  for glassy systems, and we reduce the technically more difficult
  part of the Talagrand's analysis to an explicit evaluation of the
  solution of a variational problem.

\end{abstract}
\maketitle

\section{Introduction}

The mathematical analysis
of the low temperature mean field spin glass phase has seen
enormous progresses  in the last few years,
 after more than twenty years since the
proposal of the Parisi ansatz \cite{PAR}, that led to its physical
understanding \cite{MPV}.

The first important progress was achieved by Guerra
and Toninelli \cite{GUERRATON}, who, through an interpolation
among systems of different size provided a
temperature independent proof of the existence
of the thermodynamic limit for the free-energy and other
thermodynamic quantities in a wide class of mean field spin glass systems.

The power of the interpolating method was fully appreciated by
Guerra \cite{GUERRA}, who comparing  interacting mean field
systems with  suitable paramagnets, was able to write the
free-energy as a sum of the Parisi expression plus a remainder
term, which is manifestly positive in the SK model as well as in
$p$-spin models for even $p$. The method was subsequently
generalized to deal with diluted spin glass models
\cite{FRANZLEONE,FRALEOTON, FRATON}  and more recently to analyze
the Kac limit of finite dimensional spin glasses \cite{FT}.

The last, but definitely not the least, step towards the proof of the
Parisi free energy solution has been performed by Talagrand who,
through a highly non trivial generalization of interpolation
techniques to coupled replica systems, could finally prove the
vanishing of the remainder in the Guerra formula \cite{TAL, TAL1,
  TAL2}.

Unfortunately the elegance of the Talagrand approach is
somewhat obscured by the mathematical necessity of dealing with many
technical details in the course of the proof.

In this note we would like to present the Talagrand theorem using a
physicist's approach, emphasizing  the main ideas and connecting it
with the theory of the glassy effective potentials discussed in the
physical literature \cite{FRANZ,FRANZPAR,PARFR,cfp,ZP}.

We discuss the case of the spherical $p$-spin model, where the
saddle point equations take a particularly simple form and where it is well
known that the best Parisi solution is of the simple ``one step
replica symmetry breaking kind'' (1RSB). This allows to reduce
the technically difficult part of the proof to the explicit solution of
variational equations that we analyze numerically.

Our paper comprises six sections. In the second section we define the
model and we set the basic definitions that we use in the rest of the
paper.  In the third section we illustrate the backbone of the
Talagrand theorem, specializing it to our simple case. In the fourth
section we derive the replica expressions and evaluate numerically
suitable ``effective potential functions'' whose positivity implies
the validity of the Parisi Ansatz. We conclude the argument in section
\ref{lowerbound}, where we show that the replica expressions for the
potentials provide a lower bound for their exact values. We finally
draw our conclusions.

\section{Some definitions and the Guerra formula.}

In the spirit of discussing the results in the simplest non trivial
example, we   focus on the case of the spherical $p$-spin
model.

The model considers $N$ real spin variables $\sigma_i$, $i=1,...,N$,
 subject to a global spherical constraint $\sum_{i=1}^N
 \sigma_i^2=N$, and interacting through a $p$-body Hamiltonian:
\begin{equation}
H[\sigma]= -  \sum_{i_1<i_2< \cdots
i_p}^{1,N}\, J_{i_1 \cdots i_p} \sigma_{i_1} \cdots \sigma_{i_p} \;.
\end{equation}
 The couplings $J_{i_1 \cdots i_p}$ are i.i.d. Gaussian random
variables with zero mean and variances:
\begin{equation}
\expj{J_{i_1 \cdots i_p}^2}=\frac{ p!}{2 N^{p-1}} \;,
\end{equation}
and we are not considering the case of a non
zero external magnetic field.

The best Parisi solution is known to be of 1RSB kind with $q_0=0$. The Guerra interpolating
   Hamiltonian \cite{GUERRA} to the level of 1RSB can be written as:
\begin{eqnarray}\label{guerra}
H_t[\sigma]=\sqrt{t}
H[\sigma]-\sqrt{1-t} \sum_{i=1}^N (h_i + h_{d,i}) \sigma_i \;,
\end{eqnarray}
where the interpolating parameter $t$ runs in the interval
$[0,1]$ and  $h_i$ and $h_{d,i}$ are Gaussian i.i.d. fields with zero mean and
variances that can be written in terms of a
 parameter $q \in [0,1]$ as:
 \begin{equation}
E_h\left[h_{i}^2\right]=\frac{p}{2} q^{p-1} \qquad
 E_{h_d}\left[h_{d,i}^2\right]=\frac{p}{2} (1- q^{p-1}) \;.
\label{dh}
\end{equation}
One then defines the generalized free-energy per spin:
\begin{equation}
f_t= - \frac{1}{\beta \, N \, m} \,  E_J \ln{ E_h \left[E_{h_d}
Z_t(h_i,h_{d,i})\right]^m}
  \;,
\label{ft}
\end{equation}
where $Z_t$ indicates the partition function at inverse temperature
$\beta$ computed with the Hamiltonian $H_t$, while $m$ is a number in
the interval $(0,1]$. Notice that $f_1$ is the free-energy of the
original model, while $f_0$ is the free-energy of a simple paramagnet
which can be readily computed expressing the spherical constraint
through a Lagrange multiplier and using standard saddle point method.
This gives:
\begin{equation}\label{f0guerra}
2 f_0= - \frac{1}{\beta}\min_{\lambda\in R} \left\{\lambda - \ln{ \left(\lambda
-\frac{\beta^2 p}{2} (1-(1-m)q^{p-1} )\right)} -\frac{m-1}{ m}
\ln{\left(\lambda - \frac{\beta^2 p}{2} (1-q^{p-1}) \right) }
\right\}\;,
\end{equation}
 Formula (\ref{ft})
makes a contact between the interpolating method and the cavity
method. In fact it is well known  that (\ref{ft}) can also be
written in the form \cite{RUELLE,AIZEN,MEZ1} (and for a recent
review see also \cite{GUERRA1}):
\begin{equation}\label{aizen}
f_t= - \frac{1}{\beta \, N} \,  \mathrm{E_{J,f_\gamma}} \ln{
\sum_{\gamma=1}^\infty w_{\gamma} E_d Z_t(h_{i,\gamma},h_{i,d}))} \;,
\end{equation}
where
 $w_{\gamma}=\frac{e^{-\beta f_\gamma}}{\sum_{\gamma=1}^\infty
  e^{-\beta f_\gamma}}$ are random weights derived from
  ``free-energies'' $f_\gamma$ that realize a Poisson point process with
 exponential  density $\rho(f)=\beta me^{\beta m f}$\footnote{This means that the numbers of levels in an interval of free energies
$[f_1,f_2]$ follows a Poissonian distribution with average
  ${e^{\beta m f_2}-e^{\beta m f_1}}$ and that the number of levels in
 disjont intervals are independent.},
 and the $h_{i,\gamma}$  are drawn
  independently for each $i$ and $\gamma$ from the Gaussian distributions
 defined by (\ref{dh}).   The fields $h_{i,d}$, $h_{i}$ receive then
 naturally the interpretation of cavity random fields, whose variances
 have to be determined self-consistently in order to estimate the
 free-energy $f_1$.

By simple integration by part, it is possible to see that the
free-energy $f_t$ can be written as \cite{GUERRA}:
\begin{eqnarray}\label{ftguerra}
f_t=f_0+t \frac{\beta(p-1)}{4}(1-(1-m)q^p)+\int_0^t ds\; {\cal R}_s. \,
\end{eqnarray}
The first two terms together in
(\ref{ftguerra}) provide the 1RSB replica free-energy
$f^{rep}_t$ for the interpolating model. $\mathcal{R}_t$ is the $t$
derivative of a remainder term which can be written as:
\begin{equation}\label{rem}
\mathcal{R}_t =\frac{\beta}{4}\,
(1-m)\average{q(\sigma,\tau)^p -
p q(\sigma,\tau) q^{p-1}+  (p-1) q^p }_{1,t}
+\frac{\beta}{4} m\average{q(\sigma,\tau)^p }_{0,t} \;,
\end{equation}
where $q(\sigma,\tau)$ denotes the overlap function:
\begin{eqnarray}
q(\sigma,\tau)=\frac 1 N \sum_{i=1}^N \sigma_i \tau_i \, .
\end{eqnarray}
The two averages  $\langle\cdot\rangle_{1,t}$ and
$\langle\cdot\rangle_{0,t}$ are defined  introducing two kinds of
interpolating Hamiltonians for two copies of the system with the same interactions \cite{TAL1}, of the form respectively:
\begin{equation}
H^{(1)}_t[\sigma,\tau]=\sqrt{t} H[\sigma]+\sqrt{t} H[\tau]
-\sqrt{1-t} \sum_{i=1}^N \left((h_i + h^1_{d,i}) \sigma_i + (h_i +
h^2_{d,i}) \tau_i \right) \, ,
\end{equation}
with  fields   $h_i$ and $h^r_{i,d}$, $r=1,2$, distributed
according to (\ref{dh}) with $h^1_{i,d}$ and $h^2_{i,d}$
independent, and:
\begin{equation}
H^{(0)}_t[\sigma,\tau]=\sqrt{t} H[\sigma]+\sqrt{t} H[\tau]
-\sqrt{1-t} \sum_{i=1}^N \left( (h^1_i + h^1_{d,i}) \sigma_i
+(h^2_i + h^2_{d,i}) \tau_i \right) \, ,
\end{equation}
where now, all the fields $h^1_{i}$ and $h^2_{i}$, $h^1_{i,d}$ and $h^2_{i,d}$ are  drawn
independently with the distribution (\ref{dh}).

To enlighten the notations
it is useful to define averaged partition functions:
\begin{eqnarray}
 \mathcal{Z}^{(0)}_t(h^1,h^2) &\equiv&
E_{h^1_d, h^2_d}\left[ Z^{(0)}_t(h^1,h^2,h^1_d,h^2_d) \right] \, , \nonumber\\
 \mathcal{Z}^{(1)}_t(h) & \equiv &
E_{h^1_d, h^2_d}\left[ Z^{(1)}_t(h,h^1_d,h^2_d) \right] \,,
\end{eqnarray}
and, for a
  given function of two spin configurations $k(\sigma,\tau)$,
averaged Boltzmann averages:
\begin{eqnarray}\label{Boltz}
\omega_1(k(\sigma,\tau))&\equiv& \frac{E_{h^1_d, h^2_d}
\int_{-\infty}^{+\infty}   \mathcal{D} \sigma \, \mathcal{D}
\tau
 k(\sigma,\tau)  \exp{(-\beta H^{(1)}_t[\sigma,\tau])} }{\mathcal{Z}^{(1)}_t(h)}\, ,\\
&& \nonumber \\
\omega_0(k(\sigma,\tau))&\equiv& \frac{E_{h^1_d, h^2_d}
\int_{-\infty}^{+\infty}   \mathcal{D} \sigma \, \mathcal{D}
\tau
 k(\sigma,\tau)  \exp{(-\beta H^{(0)}_t[\sigma,\tau])} }{\mathcal{Z}^{(0)}_t(h^1,h^2)}\, ,\\
&& \nonumber \\
\mathcal{D} \sigma &\equiv & d \sigma \delta\left(\sum_{i=1}^{N}
\sigma_i^2 -N \right)\nonumber\, .\\
\end{eqnarray}
With these definitions, we can finally write:
\begin{eqnarray}\label{averages}
\average{k(\sigma,\tau)}_{1,t}& \equiv &E_J \left(\frac{ E_h
 {  \mathcal{Z}^{(1)}_t(h)}^{\frac{m}{2}}
 \, \omega_1(k(\sigma,\tau))  }{
E_h {\mathcal{Z}^{(1)}_t(h)}^\frac{m}{2} }\right) \, ,\\
&&\nonumber\\
\average{k(\sigma,\tau)}_{0,t} & \equiv & E_J  \left(\frac{
 E_{h_1,h_2}  { \mathcal{Z}^{(0)}_t(h^1,h^2)}^{m}
  \omega_0(k(\sigma,\tau)) }{
    E_{h_1,h_2}  { \mathcal{Z}^{(0)}_t(h^1,h^2)}^m    }\right) \, .
\end{eqnarray}

It is apparent from the convexity of the function $q^p$ for even $p$
that both averages in the remainder (\ref{rem}) are non-negative \cite{GUERRA}.
Talagrand theorem consists in proving that in the thermodynamic limit,
for an appropriate temperature-dependent choice of the parameters $q,m$, both terms in
(\ref{rem}) are indeed equal to zero at all temperatures. For $t=1$
the free-energy $f_1$ is independent of $q$ and $m$. The minimization
of the remainder $\int_0^1 dt\; {\cal R}_t$ is therefore equivalent to
the maximization of the 1RSB free-energy. We will need  to consider the
$t$-dependent free-energy, for the values of $q$ and $m$
that achieve this maximization.

\section{Talagrand theorem.}

The Talagrand's analysis of the remainder can be divided in three
logical steps:
\begin{enumerate}
\item From the Taylor formula one simply observes that:
\begin{eqnarray}\label{Tay}
q(\sigma,\tau)^p - p q(\sigma,\tau) q^{p-1}+  (p-1) q^p  \leq
\frac{p(p-1)}{2} (q(\sigma,\tau) - q)^2 \, ,
\end{eqnarray}
and it is evident for $p>2$ that:
\begin{eqnarray}
 q(\sigma,\tau)^p
<q(\sigma,\tau)^2 \,;
\end{eqnarray}
to bound the remainder it is then enough to bound the averages
$\average{(q(\sigma,\tau) - q)^2 }_{1,t}$ and
$\average{q(\sigma,\tau)^2 }_{0,t}$.

\item In order to get estimates of the averages
  $\average{(q(\sigma,\tau) - q)^2 }_{1,t}$ and
  $\average{q(\sigma,\tau)^2 }_{0,t}$, one considers the $t$-dependent
  probabilities $P_{j,t}(q_C)= \langle
  \delta(q(\sigma,\tau)-q_{C})\rangle_{j,t}$ for $j=0,1$, that the
  overlap takes a value $q_C$, respectively in the ensembles defined
  by $\average{\cdot}_{1,t}$ and $\average{\cdot}_{0,t}$.  The main
  part of the theorem consists in proving that, given $\epsilon>0$,
  for each $t<1$, a $t$-dependent positive constant $C(t)$ exists such
  that, if $(q_C - q)^2 \geq C(t) (f_t-f^{rep}_t)+\epsilon$ and
  $q_{C}^2\geq C(t) (f_t-f^{rep}_t)+\epsilon $, then it exists a
  positive constant $C'(t,\epsilon)$ such that $P_{1,t}
  (q_C)<\exp(-C'(t,\epsilon) N)$ and $P_{0,t}(q_C)<\exp(-C'(t,\epsilon) N)$,
where $f_t^{rep}$ indicates the t-dependent free energy
calculated with the Parisi ansatz.

We will explicitely see in section \ref{numerical} that the constant $C'(t,\epsilon)$ turns
out to be a monotonically decreasing function of $t$, behaving as
$\epsilon (1-t)^2$ for $t$ close to $1$.

\item The estimates in the previous point imply, in the large $N$
  limit and fixing any $t_0 < 1$, the existence of a positive
  $t_0$-dependent constant $K(t_0)$ for which we can write the
  differential inequality:
\begin{eqnarray}
\frac{d(f_t-f^{rep}_t)}{dt}\le K(t_0) (f_t-f^{rep}_t) \, ,
\end{eqnarray}\label{diffineq}
 valid  for every $t\leq t_0$.
Since $f_{t=0}=f^{rep}_{t=0}$, (\ref{diffineq})
proves  $f_t \leq f^{rep}_t$  for every $t \leq t_0$, which by
continuity  can be extended to $t=1$.
\end{enumerate}

While the first and third points can be easily understood,
the proof of the second point requires several additional
steps.

\subsection{ Analysis of the probabilities $P_{j,t}(q_C)$.}
To estimate the probabilities $P_{j,t}(q_C)$ it is useful to relate
them to suitable ``potential functions'' for coupled replicas, similar
to the ones used in the physical literature to study off-equilibrium
configurations of glassy systems \cite{FRANZ,FRANZPAR,PARFR,cfp,ZP}.  The
potential functions can be defined as the difference between the
generalized free-energies of the interpolating system for the two
replica with and without an additional constraint in the mutual
overlap:
\begin{eqnarray}
V_{1,t}(q_C)&\equiv&-\frac{2}{m N \beta} \left[ E_J \ln  E_h \left(
 \mathcal{Z}^{(1)}_t(h) \omega_1(\delta(q(\sigma,\tau)-q_C)) \right)^\frac{m}{2}  -E_J
\ln  E_h {\mathcal{Z}^{(1)}_t}(h)^\frac{m}{2}  \right] \,,\label{pot1}\\
&&\nonumber\\
V_{0,t}(q_C)& \equiv &-\frac{1}{m N \beta} \left[E_J \ln E_{h_1,h_2} \left(
 \mathcal{Z}^{(0)}_t(h^1,h^2)  \omega_0(\delta(q(\sigma,\tau)-q_C)) \right)^m  -E_J
\ln  E_{h_1,h_2}  {\mathcal{Z}^{(0)}_t(h^1,h^2)  }^m  \right]
\nonumber \,.\label{pot0}
\end{eqnarray}
Notice, that, despite the dependence does not appear in the notation,
the Boltzmann averages $\omega(\cdot)$ depend in
(\ref{pot1},\ref{pot0}) on the same fields as the adjacent
partition functions $\mathcal{Z}^{(1)}$ and $\mathcal{Z}^{(0)}$
respectively.

One can then use the property (\ref{aizen}) and write:
\begin{eqnarray}
V_{j,t}(q_C)&=&
 -\frac{T}{N}E_{J,f_\gamma}\ln
\left(
  \frac{
    \sum_{\gamma=1}^\infty w_\gamma \mathcal{Z}^{j}_t(h_\gamma)
   \omega_{j}(\delta(q(\sigma,\tau)-q_C))
  }
{\sum_{\gamma=1}^\infty w_\gamma \mathcal{Z}^{j}_t(h_\gamma)}
\right) \,
\label{23},\\
&&\nonumber\\
P_{j,t}(q_C)&=& E_{J,f_\gamma}
\left(
  \frac{
    \sum_{\gamma=1}^\infty w_\gamma \mathcal{Z}^{j}_t(h_\gamma)
 \omega_{j}(\delta(q(\sigma,\tau)-q_C))
  }
{\sum_{\gamma=1}^\infty w_\gamma \mathcal{Z}^{j}_t(h_\gamma) } \right) \, .
\label{24}
\end{eqnarray}
Again, in formulae (\ref{23},\ref{24}) it should be
  understood that the fields appearing implicitly in the
  $\omega(\cdot)$ are the same of the adjacent partition
  functions  $\mathcal{Z}^{j}_t$.
In the weights $w_\gamma=\frac{e^{-\beta f_\gamma}}{\sum_\gamma e^{-\beta
    f_\gamma}}$, the variables $f_\gamma$ are chosen with exponential densities respectively $\rho_1(f)=e^{\frac{\beta m}{2} f}$
in the case labeled by 1, $\rho_0(f)=e^{\beta m f}$ in the case 0.

One can now immediately  see that $P_{j,t}(q_C)$   and $V_{j,t}(q_C)$ are
respectively the expectation value
and a properly normalized expectation of the logarithm  of the same
 random variable $X_{j,t}(q_C)$,
 where the averages are performed over the random couplings and
 the free-energies $f_\gamma$:
\begin{eqnarray}
V_{j,t}(q_C)&=&
 -\frac{T}{N} E_{J,f_\gamma}\ln X_{j,t}(q_C) \, ; \\
P_{j,t}(q_C)&=& E_{J,f_\gamma}  X_{j,t}(q_C) \, .
\end{eqnarray}
 It is possible to show, using techniques explained for example in
\cite{TALAGRAND}, that $1/N \ln(X_{j,t}(q_C))$ is a self-averaging quantity
and the following ``concentration of measure property'' holds:
\begin{eqnarray}
Prob\left[ X_{j,t}(q_C)>e^{-N \beta
V_{j,t}(q_C)+N\epsilon_1}\right]+ Prob\left[ X_{j,t}(q_C)<e^{-N
\beta V_{j,t}(q_C)-N \epsilon_1}\right] < 2
e^{-\frac{N\epsilon_12}{2\beta}}
\end{eqnarray}
for all $\epsilon_1>0$.  Noting that $X_{j,t}(q)$ is a probability
density, integrating it in $q$ on a small interval $[q_C,q_C+\Delta
q]$ and passing to the limit $\Delta q\to 0$, from a direct
computation one obtains:
\begin{eqnarray}
P_{j,t}(q_C) \leq 4 e^{-\frac{N \epsilon_12}{2 \beta}} +
 e^{-N \beta V_{j,t}(q_C) - \epsilon_1 N} +
e^{-N \beta V_{j,t}(q_C) +\epsilon_1 N}  \, .
\end{eqnarray}
We thus see  that
$P_{j,t}(q_C)$ is bounded by an  exponentially small number in $N$
whenever $V_{j,t}(q_C)$ is strictly positive in the limit $N\to \infty$.

The remaining part of the theorem is then devoted to showing the
positivity of the two potentials $V_{j,t}(q_C)$ for every $t<1$.  To
this end one would like to have a lower bound provided by the replica
expressions \footnote{Talagrand obtained the bound within the
  interpolating method. Some of the solutions he uses were previously
  derived in \cite{FRANZ} with the replica method.}, that are the
quantities that we are able to evaluate.  Unfortunately, with the
interpolation method we cannot directly prove the desired inequality
$V_{j,t}(q_C) \geq V_{j,t}^{rep}(q_C)$, where $V_{j,t}^{rep}(q_C)$,
$j=1,0$, indicate the replica expression of the two potentials. One
can however show that the replica expression gives a lower bound for
the generalized free energies $f_{j,t}(q_C)=V_{j,t}(q_C)+2 f_t$.
To infer from this bound the lower bound for the potentials Talagrand
found an ingenious shortcut: one first prove
the existence for all $t<1$ of a positive constant $C(t)$ such that,
for all $q_C\ne q$ one has $V_{1,t}^{rep}(q_C) >\frac{2 (q_C -q )^2}{
  C(t)}$; for $q_C$ such that $(q_C -q )^2 > C(t) (f_t -f_t^{rep})$
one can then write the following series of inequalities:

\begin{equation}
V_{1,t}(q_C)>V_{1,t}^{rep}(q_C)+ 2 f_t^{rep}(q_C) - 2 f_t>
V_{1,t}^{rep}(q_C) -\frac{2 (q_C -q )^2}{ C(t)} >0 \,.
\end{equation}
An analogous series of inequalities can be written for $V_{0,t}(q_C)$
for all $q_C\ne 0$ where one finds $V_{0,t}^{rep}(q_C) >\frac{2 q_C^2}{
  C(t)}$ for some $C(t)$.

In the next section
we  use the standard replica approach to evaluate the two potentials
$V_{j,t}^{rep}(q_C)$.

\section{The potential functions in the replica approach.}\label{repsec}

In order to  get quick heuristic estimates of $V_1$ and $V_0$  we
resort to the replica analysis, postponing to section
\ref{lowerbound} the discussion on how the expressions we find provide
free-energy lower bounds. The readers not interested to the details 
of the replica derivation can jump directly to section
\ref{numerical}.

Before entering  the discussion of the potential functions, we
find it useful to illustrate the analysis of the interpolating model
for a single copy of the system through the replica method.  This
can be done starting from expression (\ref{ft}), considering $m$
as an integer and substituting the log with an $n/m$-th power
(supposed to be an integer):
\begin{equation}\label{ftrep}
f_t= - \frac{1}{\beta \, N \, m} \;  E_J \ln{ E_h \left[E_{h_d}
Z_t(h_i,h_{d,i})\right]^m}  \rightarrow
 -
\frac{1}{\beta \, N \, n} \ln E_{J} \left[E_h\left[E_{h_d} Z_t(h,h_{d})
\right]^m \right]^{\frac{n}{m} }\;.
\end{equation}
Expanding the powers one sees that we are in presence of $n$
replicas $\sigma^a$ of the original systems divided in $n/m$
groups, each with  $m$ replicas. Together with the spin configurations
the fields $h$ are also replicated. Replicas $a$ and $b$ in the
same group have identical fields $h^a_i=h_i^b$, while replicas
$a,b$ in different groups have statistically independent fields
$h_i^a$ and $h_i^b$. The Gaussian distribution of the fields is
summarized in the covariance matrix:
\begin{eqnarray}
E[h_i^a h_i^b]+ E[h_{i,d}^a h_{i,d}^b]=H_{ab}=
\left\{\begin{tabular}{cc}
p/2 & a=b\\
p/2 $q^{p-1}$  &\; a,b \, in \,  the\, same\, group\, of \, $m$ \, replicas \\
0 & a,b\; in \; different \;\; groups
\end{tabular}
\right.
\end{eqnarray}
At this point the calculation follows the usual rails of replica
analysis of mean-field spin glasses\cite{MPV}. Introducing the
overlap matrix $Q_{ab}$ one finds:
\begin{equation}\label{ftreplica}
f^{rep}_t =- \lim_{n\to 0}\frac{1}{n} {S.P.} \left\{ \frac{\beta t}{4}  \sum_{a b}
Q_{a b}^p +
 \frac{ (1-t) \beta}{2}
\sum_{a \neq b} H_{a b} Q_{a b} + \frac{1}{2 \beta} \ln \det{ Q}
\right\}. \;
\end{equation}
where $S.P.$ denotes saddle point evaluation over the parameters
$Q_{ab}$.
The saddle point equations for $Q$ read:
\begin{eqnarray}
\frac{p\beta t}{4} Q_{ab}^{p-1}+\frac{1-t}{2}\beta H_{ab}+\frac{1}{2\beta}
(Q^{-1})_{ab} =0.
\label{spq}
\end{eqnarray}
The values of $H_{ab}$ that make
 the saddle point equations
(\ref{spq}) independent of $t$, are such that:
\begin{eqnarray}
H_{ab}=\frac{p}{2} Q_{ab}^{p-1}\,; \label{h}\\
 \beta^2\frac{p}{2} Q_{ab}^{p-1}=-(Q^{-1})_{ab}\, .
\label{spq2}
\end{eqnarray}
This is the choice needed to minimize the remainder (\ref{rem}).

The further analysis of expression (\ref{spq2}) is standard
\cite{CRIS} and will not be reproduced here.

We pass now to the slightly more involved expressions
for the potentials $V_{j,t}(q_C)$.
We  give full details for the potential $V_{1,t}(q_C)$.  An
analogous procedure can be followed for $V_{0,t}(q_C)$.
The expression for the ``constrained free
energy''  $f_{1,t}(q_C)$ is given from
(\ref{pot1}) by:
\begin{eqnarray}
&&f_{1,t}(q_C)=-\frac{2}{m N \beta}  E_J \ln  E_h \left(
\mathcal{Z}^{(1)}_t(h) \omega_1(\delta(q(\sigma,\tau)-q_C))
  \right)^\frac{m}{2}
  \qquad r=1,2 \:
 \: a=1,\dots,n \\
&&\to -\frac{2}{n N \beta}\ln
E_{J,h^{ra},h^{ra}_d} \int_{-\infty}^{+\infty} \prod_{r,a}\mathcal{D}
\sigma^r_a \,
 E_{h^{ra}_d} \exp{\left[-\beta \sum_{r,a} H_{1}(h^{ra}_d,h^{ra})\right]}
\prod_{a}\delta(q(\sigma^1_a,\sigma^2_a)-q_C) \nonumber  \,.
\end{eqnarray}
We now consider two copies, indexed by $r=1,2$, replicated $n$ times.
The fields $h^{r,a}$ and $h^{s,b}$ are equal if $a$ and $b$ belong to
the same group of $\frac{m}{2}$ replica, while they are independent in
the other case for all choices of $r$ and $s$; the fields $h^{ra}_d$
are independent for each replica $ra$.  In this case we define a field
covariance matrix ${\cal H}_{ra,sb}=E[h_i^{ra} h_i^{sb}]+
E[h_{i,d}^{ra} h_{i,d}^{sb}]$ which can be conveniently recast in the
form $ H_{a,b} ={\cal H}_{ra,rb}$ ($r=1,2$), $ \Delta_{a,b}={\cal
  H}_{ra,sb}$ ($r\ne s =1,2$). The form of these matrices is given by:
\begin{eqnarray}
H_{ab}&=&\left\{\begin{tabular}{cc}
p/2 & a=b\\
p/2 $q^{p-1}$  &a,b \; in \;  the\; same\;\; group \; of \; $m/2$ \; replica\\
0 & a,b\; in \; different \;\; groups
\end{tabular}
\right.\\
\Delta_{a b}&=&\left\{\begin{tabular}{cc}
p/2 $q^{p-1}$ &\; a=b\\
p/2 $q^{p-1}$  &\; a,b \; in \;  the \; same\;\; group \; of \; $m/2$ \; replica\\
0 & a,b\; in \; different \;\; groups
\end{tabular}
\right.
\end{eqnarray}\label{333}
As for the single replica system, the best choice for the
  values of the parameters $q$ and $m$ that appear in the
 fields' variances is the one that
  maximizes the free energy of the "free" two replica system, that is, the
  interpolating system without the additional constraint
  $q_{12}=q_C$.

As explained in \cite{FRANZ} the replica computation of the
constrained free-energy is analogous to the unconstrained case,
except that one needs now to introduce a replica matrix ${\cal
Q}_{ra,sb}$ where the elements ${\cal Q}_{ra,ra}$ are fixed to the
value $q_C$ for all $r,a$. In terms of $\mathcal{Q}$ and
$\mathcal{H}$ the free-energy has the form (\ref{ftreplica}):
\begin{equation}\label{f1rep}
f^{rep}_{1,t}(q_C)= - \lim_{n\to 0} S.P. \frac{1}{n }\, \left[ \frac{t \beta  }{4}
\sum_{ra,sb}\, \mathcal{Q}^p_{ra,sb} + \frac{\beta (1 -t)}{2}
\sum_{ra,sb}\,\mathcal{H}_{ra,sb} \mathcal{Q}_{ra,sb}+\frac{1}{2
\, \beta} \mbox{Tr} \ln{\mathcal{Q}}  \right] \;.
\end{equation}
 Supposing an ansatz
such that ${\cal Q}_{1a,1b}={\cal Q}_{2a,2b}$ and ${\cal Q}_{1a,2b}={\cal
Q}_{2a,1b}$, all to be taken as Parisi matrices, it is useful to define
the submatrices $Q_{ab}={\cal Q}_{1a,1b}$ and $P_{ab}={\cal Q}_{1a,2b}$.

Before considering the explicit solution of (\ref{f1rep}), we just
mention the difference in procedure
 to calculate the
free energy $f^{rep}_{0,t}(q_C)$. In this case, by the definition
(\ref{pot0}), the $h$ fields are statistically independent for the
two copies, and within each copy $h^a=h^b$ if $a$ and $b$
belong to the same group of $m$ replica, while $h^a$ and $h^b$
are independent in the other
case. With the same meaning as before, we can then write:
\begin{eqnarray}
H_{ab}&=&\left\{\begin{tabular}{cc}
p/2 & a=b\\
p/2 $q^{p-1}$  &a,b \; in \;  same\;\; group \; of \; $m$ \; replica\\
0 & a,b\; in \; different \;\; groups
\end{tabular}
\right.\\
&&\nonumber\\
\Delta_{ab}&=& 0 \;\; \;\; \forall \, a,b \,.
\end{eqnarray}\label{000}
keeping  this difference in mind, the formal expression for  the
free energy $f^{rep}_{0,t}(q_C)$ is still given by
(\ref{f1rep}).

We now want  to find solutions for the matrix ${\cal Q}_{ra,sb}$ in
both cases.

\subsection{Analysis of the saddle point equations}

We would like now to show the existence of solutions to the
saddle point equations that have respectively $V_{1,t}^{rep}(q_C)>0$
for $q_C\ne q$ and $V_{0,t}^{rep}(q_C)>0$ for $q_C \ne 0$.

The simplest Ansatz that we can try for the matrices $Q_{a,b} $ and
$P_{a,b} $ is of 1RSB type. 
\begin{eqnarray}\label{44}
q(x)&=& \left\{\begin{array}{ccc}
q_0=0 & 0\leq x < m_1 \\
q_1 & m_1 \leq x < 1
\end{array} \right. \\
&&\nonumber\\
\label{45}
p(x)&=& \left\{\begin{array}{ccc}
p_0=0 &  0\leq x < m_1 \\
p_1 & m_1 \leq x <1
\end{array} \right. \;,
\end{eqnarray}
As we will see below, this is a valid
Ansatz giving a positive potential for the case of
$V_{0,t}^{rep}(q_C)$.

 Unfortunately, we numerically found
that the best variational solution of this
form for $f^{rep}_{1,t}(q_C)$ gives negative values of the potential
in some range of $q_C$. A
good positive solution is obtained at the level of 2RSB, which is the
form we use, solving numerically the variational problem with respect
to the parameter $m_1$, $q_1, q_2, p_1, p_2$ of the functions:
\begin{eqnarray}
q(x)&=& \left\{\begin{array}{ccc}
q_0=0 & 0\leq x < m/2 \\
q_1 & m/2 \leq x <m_1  \\
q_2 & m_1 \leq x <1
\end{array} \right.
\label{42}\\
&&\nonumber\\
p(x)&=& \left\{\begin{array}{ccc}
p_0=0 &  0\leq x < m/2 \\
p_1 & m/2 \leq x <m_1  \\
p_2 & m_1 \leq x <1
\end{array} \right.
\label{43}
\end{eqnarray}
that parameterize respectively the matrices $Q$ and $P$. With both
parametrizations (\ref{44},\ref{45}) and (\ref{42},\ref{43}), the
saddle points in eq. (\ref{f1rep}) turn out to be maxima with
respect to all the variational parameters. The resulting expression
for the potentials in terms of the variational parameters can be
easily obtained substituting (\ref{42},\ref{43}) in (\ref{f1rep});
given its length we do not reproduce it here.

\subsection{Numerical evaluation of the variational potentials}\label{numerical}

Before analyzing the potentials for generic values of $t$ it is useful to
discuss its behavior for $t=1$ \cite{FRANZPAR,PARFR,ZP}, where
$V_{1,1}$ and $V_{0,1}$ coincide.

At temperatures higher than the ``dynamical transition temperature''
\cite{KURCH} $T_d$, a single, locally quadratic absolute minimum in
$q=0$ is present, with $V_{1,1}^{rep}(0)\equiv V_{1,1}^{rep}(0) = 0$. When the
temperature is lowered below $T_d$ but is still higher than the
thermodynamical transition temperature $T_s$, besides the $q=0$,
$V_{1,1}^{rep}(0) = 0$, a second local minimum, higher than the previous one
appears for a value of $q>0$.  When $T$ reaches $T_s$ and for values
of $T<T_s$, the replica symmetry is broken, the two minima are
degenerate $V_{1,1}^{rep}(0) = V_{1,1}^{rep}(q) = 0$ and the argument of the high
overlap minimum $q$ equals to the solution of the 1RSB variational
problem at that temperature.

The behavior of the $t<1$ replica potentials is investigated solving
numerically the variational equations with respect to all the
variational parameters. 

Let us first discuss the simple case of
the high temperature replica symmetric phase for $T>T_s$.  In this
case only the $V_{0,t}^{rep}$ potential has to be considered and we show in
fig.  \ref{potRS.fig} that the required inequality
$V^{rep}_{0,t}(q_C)>0$ for $q_C \neq 0$, already satisfied for $t=1$,
remains valid for every $t<1$.

Let us then analyze the RSB phase
 ($T\leq T_s$) when $t<1$ and show that the
minimum in zero (case 1) or in $q$ (case 0) becomes respectively
higher than the absolute one.

 We show in fig. \ref{zero_1.fig} the results of the saddle point
 evaluation of the potential $V_{0,t}^{rep}(q_C)$
 for different
values of the parameter $t$ at a fixed temperature
 $T<T_s$, having verified that the  behavior does not
 change for different values of temperature below the static
 transition.

 In fig. \ref{uno_1.fig} we plot the potential $V_{1,t}^{rep}(q_C)$
 for different values of the parameter $t$
at the same temperature
 $T<T_s$ as in fig. \ref{zero_1.fig}.

The two potentials $V^{rep}_{1,t}(q_C)$ and $V^{rep}_{0,t}(q_C)$ are
greater than zero for every value of $q_C$ different,
respectively, from $q$ and zero, and are quadratic near the
(lower) minimum.

We numerically find a quadratic dependence on $1-t$, for $t\simeq 1$, of the
value of the potential in the higher local minimum (fig. \ref{zerot.fig},
\ref{unot.fig}), implying that the constant $1/C(t)$ behaves as
$(1-t)^2$ for $t$ close to one.

\section{Lower bound for the interpolating free energy.}\label{lowerbound}

 To conclude the argument  we now show
 that the constrained free-energies calculated in the previous section within
 the replica approach indeed supply lower bounds for the true ones.

We just repeat here  the argument given by Talagrand, specializing
 to our simple case. For each value of $t$ we
define an interpolating Hamiltonian $H_u$ that relates, varying
the parameter $u$ in the interval $[0,1]$, the two replica
interpolating system  with a suitable paramagnet.

Once again, we only discuss the case of
$f_{1,t}(q_C)$,  the same  procedure being valid with minor
modifications for $f_{0,t}(q_C)$.

 We have found in the previous section a good positive solution at the level of
2RSB; to deal with this case within an interpolating scheme we need three
families of fields for each copy. One  then defines:
\begin{eqnarray}
H_{u}[\sigma,\tau]&=&\sqrt{u t}
H_p[\sigma] + \sqrt{u t} H_p[\tau]-\sqrt{1-t} \sum_{i=1}^N \left( (h_i + h^1_{d,i}) \sigma_i
 + (h_i  + h^2_{d,i})   \tau_i\right)
\nonumber \\
&-&\sqrt{(1-u) t} \sum_{i=1}^N \left(  (g^1_{1,i} +g^1_{2,i} + g^1_{d,i}) \sigma_i +  (g^2_{1,i} +g^2_{2,i} + g^2_{d,i} ) \tau_i
 \right) \;,
\end{eqnarray}
where the fields $h$ and $h_d$ are defined as in (\ref{dh}) and
$h^1_d$ and $h^2_d$ chosen independently, while the variances and the
covariances of the fields $g$ can be written, in a notation consistent with
section \ref{repsec}, as:
\begin{eqnarray}\label{dg}
&& E_g\left[{(g^r_{1})}^2\right]=\frac{p}{2} q_1^{p-1} \qquad
E_g\left[{(g^r_{2})}^2\right]=\frac{p}{2} q_2^{p-1} \qquad
 E_g\left[{(g^r_{d})}^2\right]=\frac{p}{2} (1 -q_2^{p-1}) \qquad r=1,2 \nonumber \\
&& \nonumber\\
&&   E_g\left[g^1_1 g^2_1\right]=\frac{p}{2}  p_1^{p-1} \qquad  E_g\left[g^1_2
g^2_2\right]=\frac{p}{2}  p_2^{p-1} \qquad E_g\left[g^1_{d} g^2_{d}
\right]=\frac{p}{2} (q_C^{p-1} -p_2^{p-1}) \, .
\end{eqnarray}
In a completely analogous way as in (\ref{ft}), we define
 the free energy:
\begin{eqnarray}
\tilde{f}_u &=& -\frac{2}{\beta \, N \, m} \,  E_J
\ln\left[E_{g^1_1,g^2_1,h} \left(E_{g^1_2,g^2_2}
  {\mathcal{Z}_{C,u}}^{m_1} \right)^{\frac{m}{2 m_1}}\right] \, ,\\
  && \nonumber\\
\mathcal{Z}_{C,u}&=&   \int_{-\infty}^{+\infty}   \mathcal{D} \sigma \, \mathcal{D} \tau
 E_{h^1_d, h^2_d, g^1_d, g^2_d}  \delta(q(\sigma,\tau) -q_C) \exp -\beta H_{u}[\sigma,\tau]\,, \nonumber
\label{fu}
\end{eqnarray}
where it is evident that $\tilde{f}_{u=1,t}= f_{t}(q_C)$.  On the
other hand, one can check that $\tilde{f}_{u=0,t}$ reproduces the non
trivial part of the replica expression of $f_{t}(q_C)$ and one has
\begin{equation}\label{lowerrep}
f^{rep}_{t}(q_C)=\tilde{f}_{u=0,t}+ \frac{ t \beta (p-1)}{2}
\left\{1-(1-m_1) q_2^p-
 (m_1-\frac{m}{2}) q_1^p
+ q_C^p -  (1-m_1) p_2^p- (m_1-\frac{m}{2}) p_1^p \right\} \; ,
\end{equation}
when the parameters that appear in the variances (\ref{dg})
maximize the expression (\ref{lowerrep}) for $t=1$.

We can thus repeat the argument as for the "standard"
interpolating method: by the usual  integrations
by part one sees that
 the derivative $\frac{d \tilde{f}_u }{d u}$ can be written, for every $t$,
as:
\begin{equation}
\frac{d \tilde{f}_u }{d u} = \frac{ t \beta (p-1)}{2} \left\{1-(1-m_1) q_2^p-
 (m_1-\frac{m}{2}) q_1^p
+ q_C^p -  (1-m_1) p_2^p- (m_1-\frac{m}{2}) p_1^p \right\} +  \mathcal{R}_u \;,
\end{equation}
where $\mathcal{R}_u$ is a  remainder with the explicit
expression:
\begin{eqnarray}
\mathcal{R}_u&=&\frac{t \beta}{2} (1-m_1) \average{2 q(\sigma,\tau )^p - p q(\sigma,\tau)
(q_2^{p-1} + p_2^{p-1} - q_1^{p-1} - p_1^{p-1}) + (p-1)
(q_2^{p} + p_2^{p} - q_1^{p} - p_1^{p}) }_{2,u} \nonumber \\
& +&\frac{t \beta}{2} (m_1 -\frac{m}{2})
 \average{2 q(\sigma,\tau )^p - p q(\sigma,\tau)
(q_1^{p-1} + p_1^{p-1}) + (p-1)
(q_1^{p} + p_1^{p}) }_{1,u} +t \beta \frac{m}{2} \average{q(\sigma,\tau )^p}_{0,u} \;.
\end{eqnarray}
The inequality $\tilde{f}_{u=0} =
f^{rep}_{1,t}(q_C) \leq f_{1,t}(q_C) = \tilde{f}_{u=1}$ follows from the
positivity of  $\mathcal{R}_u$, that is evident from the convexity
of the function $q^p$ for even $p$, in the same way as in
(\ref{rem}). In the Appendix we give the expressions for the
averages that appear in the remainder.

\section{Conclusion.}

The main point of this paper is to present the Talagrand proof of the
Parisi anzatz in disordered mean field models in the physicist's
language.  With the aim of emphasizing the conceptual aspects of the
proof, we have specialized it to the case of the spherical $p$-spin
model, where the simplicity of the Parisi solution allows a great
reduction of technical difficulties.  The main advantages of our
approach  is that we circumvent the most involved part of the
demonstration by a direct solution of a variational problem, which in
the case faced in this paper allows for an easy numerical solution.
In doing that, we emphasize the connections of the Talagrand proof
with the well known method of coupled replicas and the construction of
glassy potential functions previously known in the physical literature
\cite{FRANZ}.  Though our paper does not really contain new
mathematical results, we hope that it could be useful to spread the
Talagrand's results in a larger community.

One of the difficulties of the Talagrand paper is that the  proof was provided in a general Parisi scheme
with arbitrary steps of RSB. This is needed in the case of the SK
model or in the low temperature phase of the Ising $p$-spin model.
There, the derivation is complicated by the need to go to the limit of
continuous replica symmetry breaking. In particular, Talagrand showed
that, for every $t<1$, it is possible to consider a number $k=k(t)$ of
steps of RSB such that an interpolating system with $j>k$ steps is
solved by the Parisi Ansatz with $j$ steps. One then has to show the
positivity of $j$ different potential functions.  An alternative is to
work directly in the continuous RSB limit, and to look for a lower
bound for a suitable family of potential functions depending on a
continuous index. These potentials can be computed explicitly in spherical
models with full RSB \cite{NIEUW},
and we hope to report soon on that case.

Thanks to the more physical approach followed in this revisitation
of the Talagrand proof, it can also be easier to analyze  more
involved mean field systems, like those diluted, where a lower
bound for the free energy was recently found \cite{FRANZLEONE,
FRALEOTON} by an extension of the interpolating procedure.

\section{Appendix.}
Generalizing the procedure used to define the averages
(\ref{averages}), the first step to define
 $\average{\cdot}_{0,u}$, $\average{\cdot}_{1,u}$ and $\average{\cdot}_{2,u}$
 is to consider three four-replica Hamiltonian with the general form:
\begin{eqnarray}
\mathbf{H}&=&\sqrt{u t} H_p[\sigma] + \sqrt{u t}
H_p[\tau]-\sqrt{1-t} \sum_{i=1}^N \left( (h_i + h^1_{d,i})
\sigma_i
 + (h_i  + h^2_{d,i})   \tau_i\right)
\nonumber \\
&-&\sqrt{(1-u) t} \sum_{i=1}^N \left(  (g^1_{1,i} +g^1_{2,i} +
g^1_{d,i}) \sigma_i + (g^2_{i,1} +g^2_{i,2} + g^2_{d,i} ) \tau_i
 \right) \nonumber \\
&+&\sqrt{u t} H_p[\sigma'] + \sqrt{u t} H_p[\tau']-\sqrt{1-t}
\sum_{i=1}^N \left( (\tilde{h}_i + \tilde{h}^1_{d,i}) \sigma'_i
 + (\tilde{h}_i  + \tilde{h}^2_{d,i})   \tau'_i\right)
\nonumber \\
&-&\sqrt{(1-u) t} \sum_{i=1}^N \left(  (\tilde{g}^1_{1,i}
+\tilde{g}^1_{2,i} + \tilde{g}^1_{d,i}) \sigma'_i +
(\tilde{g}^2_{i,1} +\tilde{g}^2_{i,2} + \tilde{g}^2_{d,i} )
\tau'_i
 \right) \;,
\end{eqnarray}
where the new introduced fields  have the same distribution of
the corresponding $h$ and $g$ fields: $\tilde{h}$ have the same
 distribution of
$h$, and $\tilde{g}$  the same distribution of $g$.

 The three different  Hamiltonians are specified by:
\begin{enumerate}
\item $\mathbf{H_0}$:  the $\tilde{h}$ and $\tilde{g}$ fields
 are
 respectively independent of the fields $h$ and $g$.

\item $\mathbf{H_1}$: $\tilde{g}^1_1=g^1_1$,
$\tilde{g}^2_1=g^2_1$ and $\tilde{h}=h$, while the
other $\tilde{h}$ and $\tilde{g}$ fields
 are
independent of the corresponding $h$ and $g$ fields.

\item $\mathbf{H_2}$: $\tilde{g}^1_1=g^1_1$,
$\tilde{g}^2_1=g^2_1$, $\tilde{h}=h$, $\tilde{g}^1_2=g^1_2$,
$\tilde{g}^2_2=g^2_2$; the "annealed" fields $h^r_{i,d}$,
 $\tilde{h}^r_{i,d}$,  $g^r_{i,d}$ and
$\tilde{g}^r_{i,d}$  ($r=1,2$)
 are always independent.

\end{enumerate}
We can now define, as done in (\ref{Boltz}), the
   Boltzmann averages of a given function
 of two spin configurations $k(\sigma,\tau)$, averaged
 over all the annealed fields
$h^r_{i,d}$, $\tilde{h}^r_{i,d}$,  $g^r_{i,d}$,
$\tilde{g}^r_{i,d}$, $r=1,2$:
\begin{eqnarray}\label{Boltz2}
\Omega(k(\sigma,\tau))_{2,1,0}&\equiv&
   \frac{\int_{-\infty}^{+\infty}E_{h^r_{i,d},\tilde{h}^r_{i,d},g^r_{i,d},\tilde{g}^r_{i,d}}
   \mathcal{D} \sigma \, \mathcal{D} \tau
 k(\sigma,\tau)
  \exp{(-\beta \mathbf{H_{2,1,0}})}}{\mathcal{Z}_{2,1,0}} \qquad r=1,2 \, ,\\
&& \nonumber \\
 \mathcal{Z}_{2,1,0}& \equiv&
\int_{-\infty}^{+\infty} E_{h^r_{i,d},\tilde{h}^r_{i,d},g^r_{i,d},\tilde{g}^r_{i,d}}  \mathcal{D} \sigma \, \mathcal{D} \tau
 \delta(q(\sigma,\sigma')-q_C) \delta(q(\tau,\tau')-q_C)  \exp{(-\beta \mathbf{H_{2,1,0}})}  \,.\nonumber
\end{eqnarray}

 The three averages that appear in the remainder
are then defined by:
\begin{eqnarray}
\average{k(\sigma,\tau)}_{2,u}&\equiv &E_J   \frac{  E_{g^1_1,g^2_1,h}
\left[
 \left( E_{g^1_2,g^2_2} \mathcal{Z}_{2}^{\frac{m_1}{2}} \right)^{\frac{m}{2 m_1} - 1}
   E_{g^1_2, g^2_2} {\mathcal{Z}_{2}}^\frac{m_1}{2}
 \Omega(k(\sigma,\tau))_{2}   \right] }{
 E_{g^1_1,g^2_1,h}
 \left( E_{g^1_2,g^2_2} {\mathcal{Z}_{2}}^{\frac{m_1}{2}} \right)^{\frac{m}{2
 m_1}} }  \, ;\\
&&\nonumber\\
\average{k(\sigma,\tau)}_{1,u}&\equiv&E_J   \frac{  E_{g^1_1,g^2_1,h}
\left[
 \left( E_{g^1_2,g^2_2,\tilde{g}^1_2,\tilde{g}^2_2} {\mathcal{Z}_{1}}^{m_1} \right)^{\frac{m}{4 m_1} - 1}
   E_{g^1_2, g^2_2,\tilde{g}^1_2,\tilde{g}^2_2} {{\mathcal{Z}_{1}}}^{m_1}
 {\Omega(k(\sigma,\tau))}_{1}   \right] }{
  E_{g^1_1,g^2_1,h}
 \left( E_{g^1_2,g^2_2,\tilde{g}^1_2,\tilde{g}^2_2} {\mathcal{Z}_{1}}^{m_1} \right)^{\frac{m}{4 m_1} }  } \, ;  \\
&&\nonumber\\
\average{k(\sigma,\tau)}_{0,u}&\equiv&E_J   \frac{
E_{g^1_1,g^2_1,h,\tilde{g}^1_1,\tilde{g}^2_1,\tilde{h}} \left[
 \left( E_{g^1_2,g^2_2,\tilde{g}^1_2,\tilde{g}^2_2} {\mathcal{Z}_{0}}^{m_1} \right)^{\frac{m}{2 m_1} - 2}
   E_{g^1_2, g^2_2,\tilde{g}^1_2,\tilde{g}^2_2} {{\mathcal{Z}_{0}}}^{m_1}
  {\Omega(k(\sigma,\tau))}_{0}   \right] }{
 E_{g^1_1,g^2_1,h,\tilde{g}^1_1,\tilde{g}^2_1,\tilde{h}}
 \left( E_{g^1_2,g^2_2,\tilde{g}^1_2,\tilde{g}^2_2}
  {\mathcal{Z}_{0}}^{m_1} \right)^{\frac{m}{2 m_1}}     } \, .
\end{eqnarray}

\newpage

\begin{center}{\bf Figure captions}
\end{center}

Figure 1: The potential $V^{rep}_{0,t}(q_C)$ for different values of $t$
at the temperature $ T_s=0.503 <T=0.526 <T_d=0.544 $, for $p=4$. From the
bottom to the top the parameter $t$ runs from 1 to 0, decreasing
by 0.1 at each plot. We have plotted only the results for $q_C\geq
0$, being the potential  symmetric for negative values of $q_C$.

Figure 2: The potential $V^{rep}_{1,t}(q_C)$ at the inverse
temperature $\beta=3$ and $p=4$. From the bottom to the top the parameter
$t$ runs from 1 to 0.9, decreasing by 0.01 at each plot. In the
zoom is shown the potential near the higher minimum at the same
inverse temperature and for the same values of the parameter $t$.
  Note that the potential is not symmetric
with respect to positive and negative values of $q_C$. 

Figure 3: 
The potential $V^{rep}_{0,t}(q_C)$ at the inverse
temperature $\beta=3$ and $p=4$. From the bottom to the top the parameter
$t$ runs from 1 to 0, decreasing by 0.1 at each plot. In the zoom
is shown the potential at the same inverse temperature near the
higher minimum, for $t$ varying from 1 (at the bottom) to 0.9 (at
the top) decreasing by 0.01 at each plot.   Note that the
potential is now  symmetric with respect to positive and negative
values of $q_C$, being  the minimum at $q_{12}=0$.

Figure 4: Value of the potential $V^{rep}_{1,t}$
 in the higher local minimum, as a function
of $1-t$, at the inverse
temperature $\beta=3$ and $p=4$. The continuous line is a quadratic fit with the function
$f(x)=a x2$, $a=2.26684$.

Figure 5: Value of the potential $V^{rep}_{0,t}$
 in the higher local minimum, as a function
of $1-t$, at the inverse
temperature $\beta=3$ and $p=4$. The continuous line is a quadratic fit with the function
$f(x)=a x^2$, $ a=0.149782$.

\begin{figure}
\includegraphics[width=14cm]{pspinRS.eps}
\caption{}
\label{potRS.fig}
\end{figure}

\begin{figure}
\includegraphics[width=14cm]{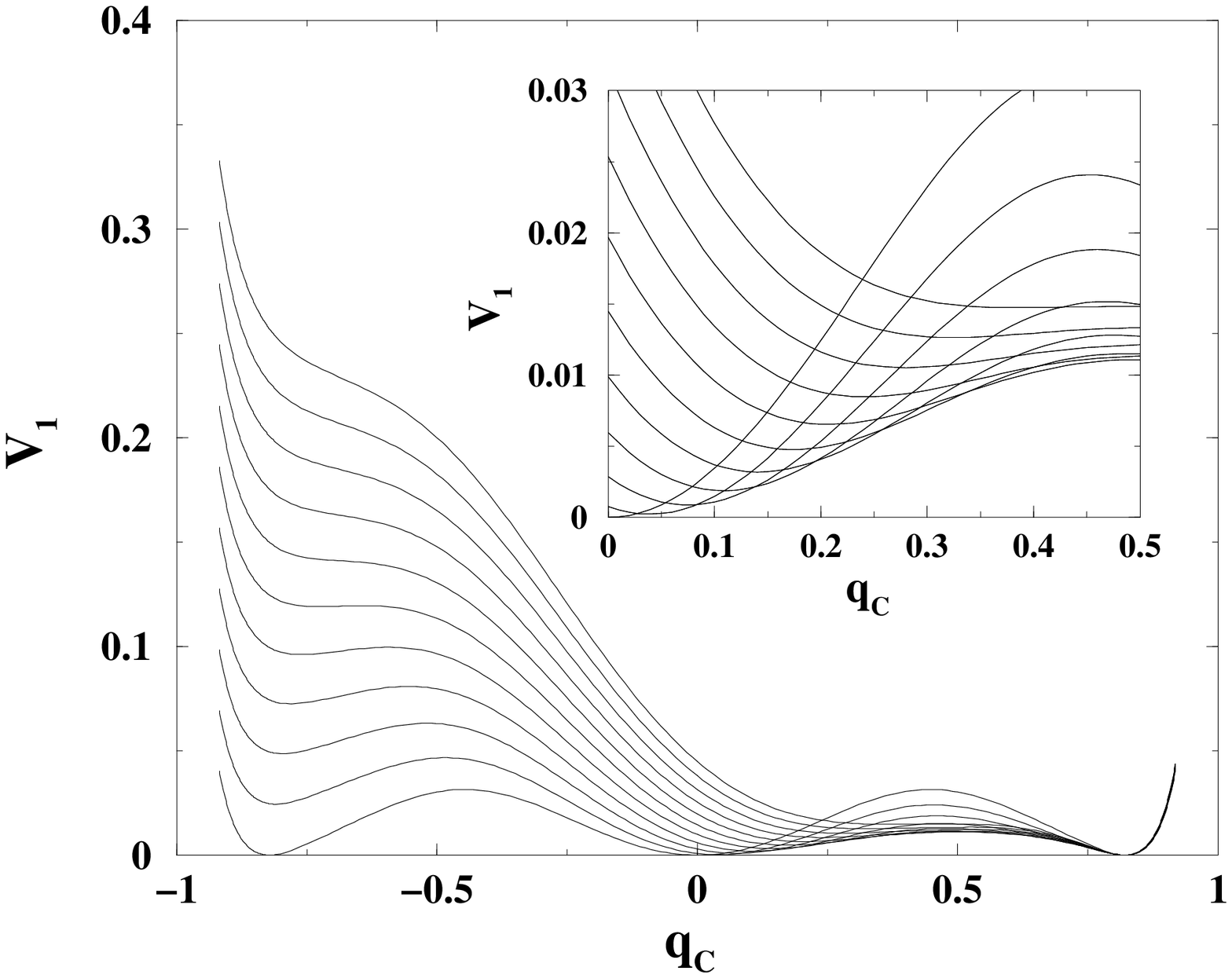}
\caption{}
\label{uno_1.fig}
\end{figure}

\begin{figure}
\includegraphics[width=14cm]{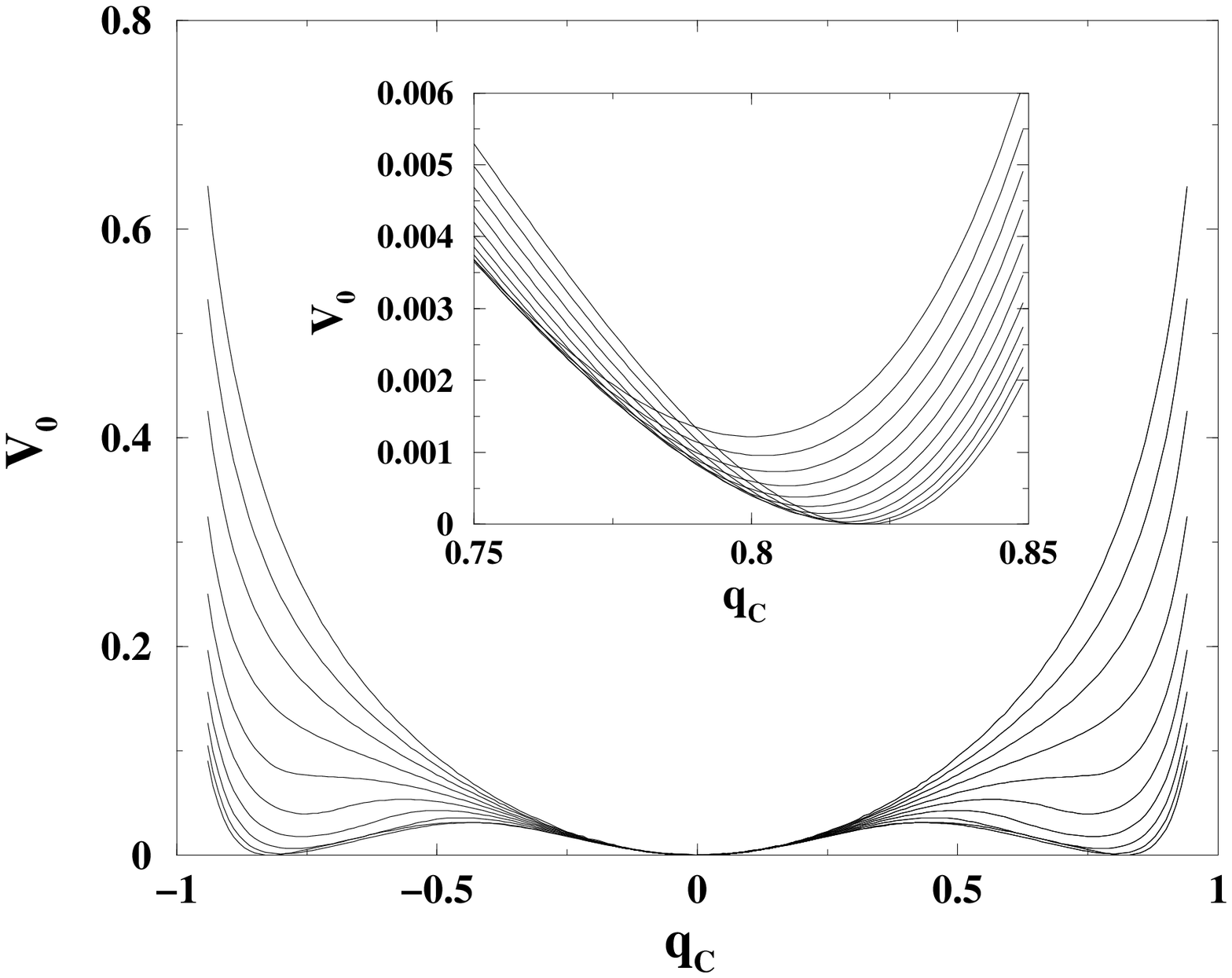}
\caption{}
\label{zero_1.fig}
\end{figure}

\begin{figure}
\includegraphics[width=14cm]{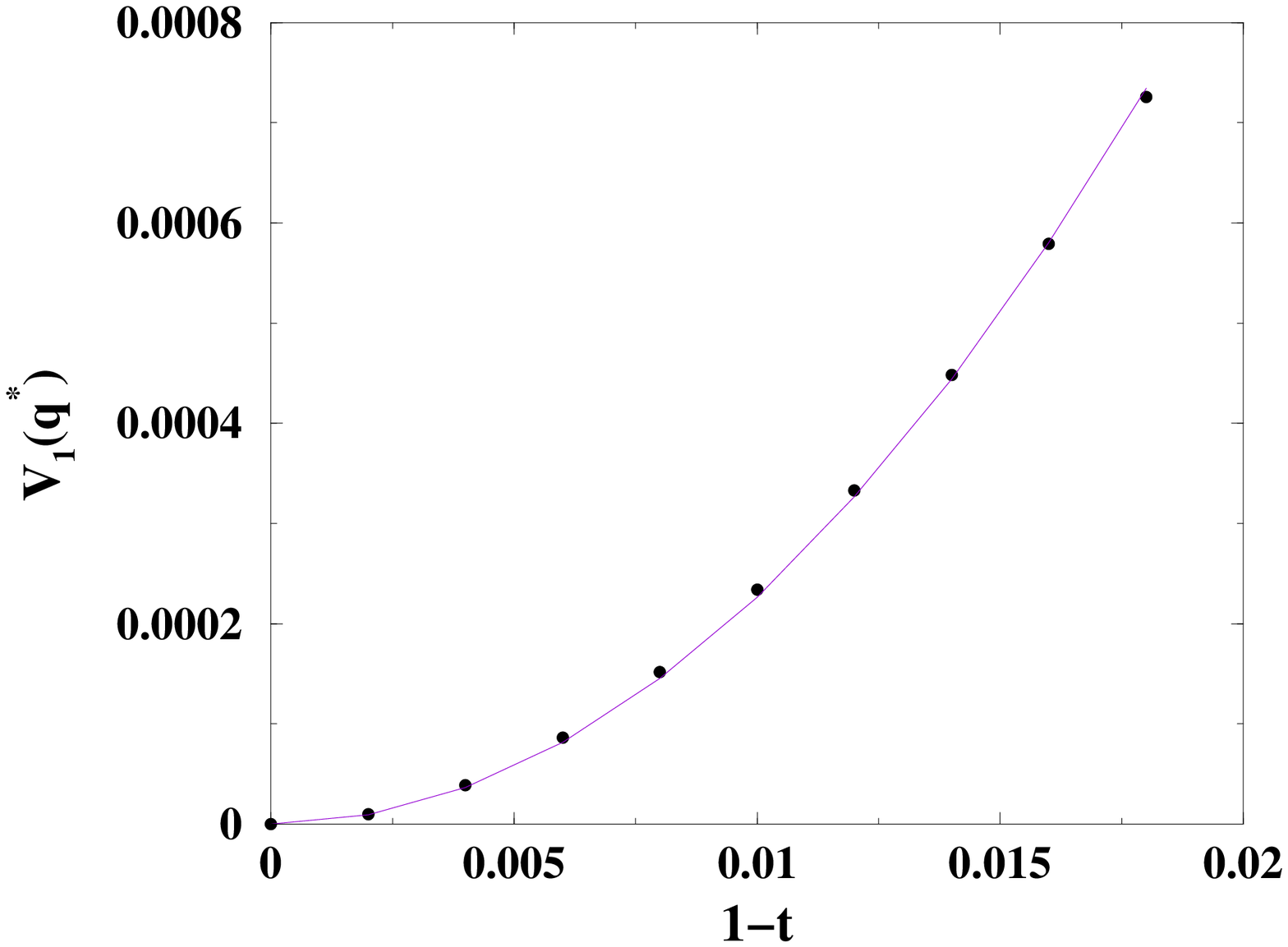}
\caption{}
\label{unot.fig}
\end{figure}

\begin{figure}
\includegraphics[width=14cm]{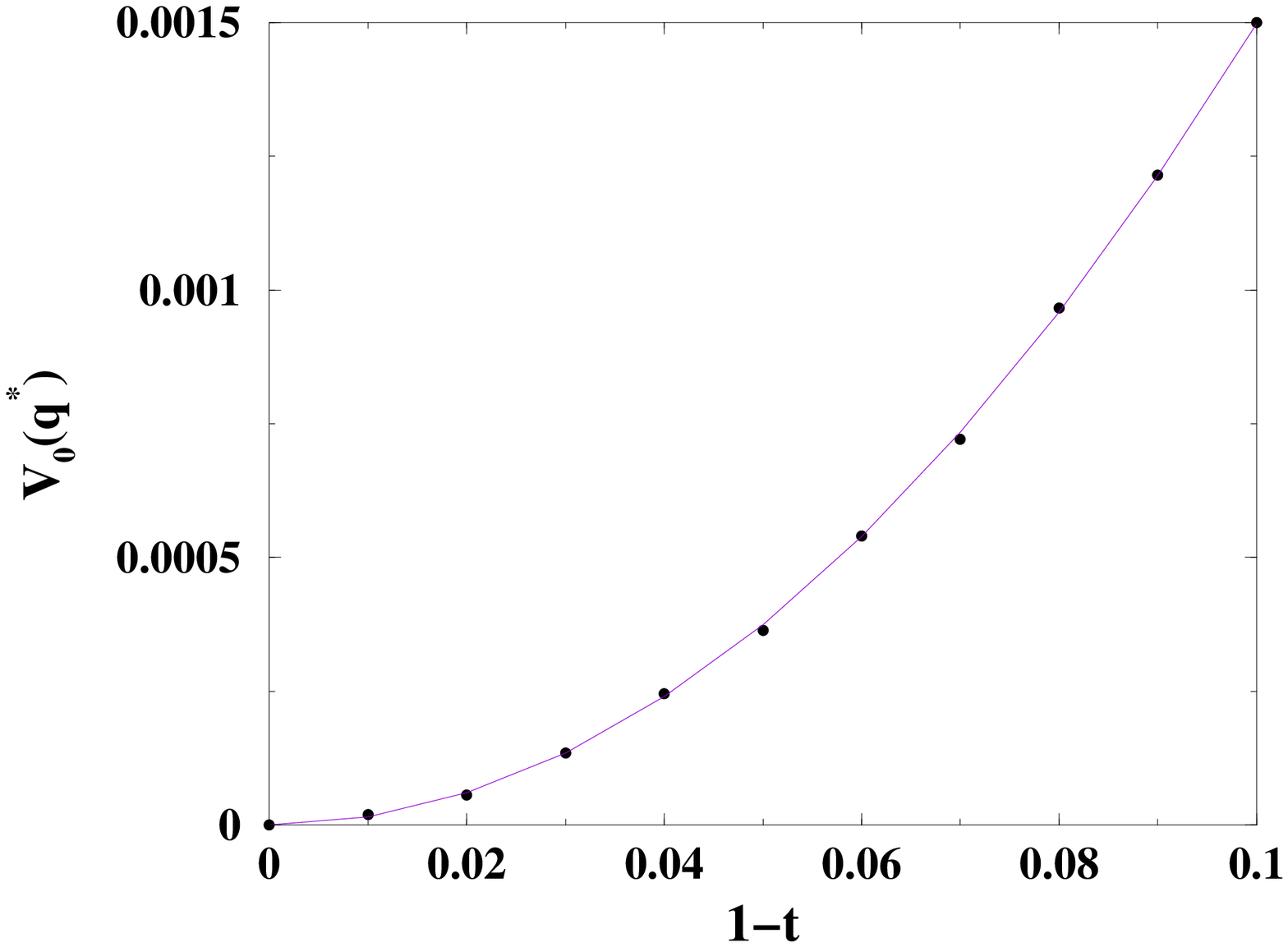}
\caption{}
\label{zerot.fig}
\end{figure}

\end{document}